\documentclass{PoS}

\title{Constraints on non-standard flavor-dependent interactions from Superkamiokande and Hyperkamiokande}

\ShortTitle{Constraints on non-standard flavor-dependent interactions from Superkamiokande and Hyperkamiokande}

\author{\speaker{OSAMU YASUDA}\\
        Department of Physics, Tokyo Metropolitan University\\
Hachioji, Tokyo 192-0397, Japan\\
        E-mail: \email{yasuda AT phys.se.tmu.ac.jp}}

\abstract{
We investigate the constraint on the flavor-dependent neutral current
Non-Standard Interactions in propagation from atmospheric neutrino
experiments Superkamiokande and Hyperkamiokande.
With the ansatz where the parameters which have
strong constraints from other experiments are neglected, we show how these
experiments put constraints on the remaining parameters of the Non-Standard
Interactions.}

\FullConference{16th International Workshop on Neutrino Factories and Future Neutrino Beam Facilities\\
                 25 -30 August, 2014\\
                 University of Glasgow, United Kingdom}

\begin{document}

\section{Introduction}
The phenomenon of neutrino oscillation is described by
the mixing matrix
\begin{eqnarray}
U=\left(
\begin{array}{ccc}
c_{12}c_{13} & s_{12}c_{13} &  s_{13}e^{-i\delta}\nonumber\\
-s_{12}c_{23}-c_{12}s_{23}s_{13}e^{i\delta} & 
c_{12}c_{23}-s_{12}s_{23}s_{13}e^{i\delta} & s_{23}c_{13}\nonumber\\
s_{12}s_{23}-c_{12}c_{23}s_{13}e^{i\delta} & 
-c_{12}s_{23}-s_{12}c_{23}s_{13}e^{i\delta} & c_{23}c_{13}\nonumber\\
\end{array}\right),
\label{eqn:mns}
\end{eqnarray}
where $s_{jk}\equiv\sin\theta_{jk}$ and
$c_{jk}\equiv\cos\theta_{jk}$, $\theta_{jk}$ with
$(j,k)=(1,2), (1,3), (2,3)$
are the three mixing angles and $\delta$ is
the Dirac CP phase.
Thanks to recent neutrino experiments \cite{Agashe:2014kda},
all the mixing angles and the mass squared differences
have been measured, and the only unknown quantities
which can probed by neutrino oscillation
are the mass hierarchy pattern and $\delta$.
It is believed that these unknown quantities will be
determined in the future neutrino experiments,
including those with intense accelerator neutrino
beams \cite{Abe:2014oxa,Adams:2013qkq}.
It is also expected that these future experiments
with intense accelerator neutrino beams will
enable us to probe new physics beyond the
standard model with massive neutrinos,
by looking for the deviation from the standard scenario.

In the standard model with three massive neutrinos,
the Dirac equation for the flavor eigenstate
$\Psi^T\equiv(\nu_e,\nu_\mu,\nu_\tau)$ of neutrino in matter
is given by
\begin{eqnarray}
i{d\Psi \over dt}=
\left[U \mbox{\rm diag}\left(E_1,E_2,E_3\right) U^{-1}
+{\cal A}
\right]\Psi,
\label{sch1}
\end{eqnarray}
where the matter potential is given by
\begin{eqnarray}
{\cal A}=A\left(
\begin{array}{ccc}
1 & 0 & 0\\
0 & 0 & 0\\
0 & 0 & 0
\end{array}
\right).
\label{matter-std}
\end{eqnarray}
Here $A\equiv \sqrt{2} G_F n_e$
stands for the magnitude of
the standard matter effect due to the
charged current interaction,
$n_e$ is the number density of the electron in the matter,
and the matter effect due to the
neutral current interaction, which is proportional
to the unit matrix in the flavor basis, is ignored
because it would affect only the phase of the
oscillation probability amplitude.

Here I would like to consider the
flavor-dependent nonstandard four-fermi interactions
\begin{eqnarray}
{\cal L}_{\mbox{\rm\scriptsize eff}}^{\mbox{\rm\scriptsize NSI}} =
-2\sqrt{2}\, \epsilon_{\alpha\beta}^{fP} G_F
(\overline{\nu}_\alpha \gamma_\mu P_L \nu_\beta)\,
(\overline{f} \gamma^\mu P f),
\label{NSIop}
\end{eqnarray}
where only the interactions with $f = e, u, d$ are relevant to
the flavor transition of neutrino due to the matter effect,
$G_F$ denotes the Fermi coupling constant, $P$ stands for
a projection operator and is either
$P_L\equiv (1-\gamma_5)/2$ or $P_R\equiv (1+\gamma_5)/2$.
(\ref{NSIop}) is the most general form of the
interactions which conserve electric charge, color, and
lepton number~\cite{Davidson:2003ha}.
In the presence of these interactions (\ref{NSIop}),
the matter potential is modified to
\begin{eqnarray}
{\cal A} \to A\left(
\begin{array}{ccc}
1+ \epsilon_{ee} & \epsilon_{e\mu} & \epsilon_{e\tau}\\
\epsilon_{e\mu}^\ast & \epsilon_{\mu\mu} & \epsilon_{\mu\tau}\\
\epsilon_{e\tau}^\ast & \epsilon_{\mu\tau}^\ast & \epsilon_{\tau\tau}
\end{array}
\right),
\label{matter-nsi}
\end{eqnarray}
where 
$\epsilon_{\alpha\beta}$ are defined as
$\epsilon_{\alpha\beta}
\equiv \sum_{f,P}(n_f/n_e) \epsilon_{\alpha\beta}^{fP}
\simeq \sum_{P}
\left(
\epsilon_{\alpha\beta}^{eP}
+ 3 \epsilon_{\alpha\beta}^{uP}
+ 3 \epsilon_{\alpha\beta}^{dP}
\right)$,
$n_f$ is the number density of $f$ in matter,
and we have taken into account the fact that the number density of
$u$ quarks and $d$ quarks are three times as that of
electrons.
The constraint on $\epsilon_{\alpha\beta}$ haven been
given by a number of works, and can be summarized as \cite{Biggio:2009nt}
\begin{eqnarray}
\left(
\begin{array}{ccc}
|\epsilon_{ee}|< 4\times 10^0 & |\epsilon_{e\mu}| < 3\times 10^{-1}
& |\epsilon_{e\tau}| < 3\times 10^0\\
& |\epsilon_{\mu\mu}| < 7\times 10^{-2}
& |\epsilon_{\mu\tau}| < 3\times 10^{-1}\\
& & |\epsilon_{\tau\tau}| < 2 \times 10^1
\end{array}
\right).
\label{b-eps-0}
\end{eqnarray}
From Eq. (\ref{b-eps-0}) we see that
\begin{eqnarray}
\epsilon_{e\mu} \simeq \epsilon_{\mu\mu} \simeq \epsilon_{\mu\tau} \simeq 0
\label{eps-mu}
\end{eqnarray}
is satisfied.

On the other hand, it was shown \cite{Friedland:2005vy}
that the non-standard matter potential (\ref{matter-nsi})
is consistent with the high energy atmospheric neutrino data
only if
\begin{eqnarray}
\epsilon_{\tau\tau} \simeq 
\frac{|\epsilon_{e\tau}|^2}
{1 + \epsilon_{ee}}
\label{b-eps-3}
\end{eqnarray}
is satisfied.

In this talk I will discuss the constraints
on the flavor-dependent non-standard interactions
from the Superkamiokande atmospheric neutrino data.
Taking the constraints (\ref{eps-mu}) and (\ref{b-eps-3})
into consideration, for simplicity I will take
the following ansatz for the matter potential\footnote{
The constraints on $\epsilon_{ee}$ and $\epsilon_{e\tau}$
from the atmospheric neutrino
have been discussed in
Refs~\cite{GonzalezGarcia:2011my, Mitsuka:2011ty, Gonzalez-Garcia:2013usa}
with the ansatz different from ours}:
\begin{eqnarray}
A\left(
\begin{array}{ccc}
1+ \epsilon_{ee} & 0 & \epsilon_{e\tau}\\
0 & 0 & 0\\
\epsilon_{e\tau}^\ast &0 & \frac{|\epsilon_{e\tau}|^2}
{1 + \epsilon_{ee}}
\end{array}
\right).
\label{matter-nsi1}
\end{eqnarray}
I will also give the expected sensitivity to the
same parameters from the future atmospheric neutrino
data by the Hyperkamiokande experiment\,\cite{Abe:2011ts}.
The result will be given as the allowed region in
the ($\epsilon_{ee}$, $|\epsilon_{e\tau}|$) plane
by marginalizing the $\chi^2$ with respect to
the standard oscillation parameters as well as
arg($\epsilon_{e\tau}$).

\section{Analysis}

The data we analyzed is for 3903 days\,\cite{Itow:2013zza}.
The analysis\,\cite{Itow:2013zza} by the Superkamiokande
collaboration uses information which is even more
detailed than the one in Ref. \cite{Ashie:2005ik}, and
we have been unable to reproduce their results of the Monte Carlo
simulation.  So we have combined the two sub-GeV $\mu$-like
data set in one, the two multi-GeV e-like in one,
the two partially contained event data set and
the multi-GeV $\mu$-like in one, and
the three upward going $\mu$ in one.
The analysis was performed with the code which were
used in Refs. \cite{Foot:1998iw,Yasuda:1998mh,Yasuda:2000de}.
$\chi^2$ is defined as
\begin{eqnarray}
\chi^2=
\min_{\theta_{23},|\Delta m^2_{32}|,\delta, \mbox{\rm arg}(\epsilon_{e\tau})}
\left[
\chi_{\rm sub-GeV}^2+\chi_{\rm multi-GeV}^2
+\chi_{\rm upward}^2\right]\,,
\label{eqn:chi}
\end{eqnarray}
where
\begin{eqnarray}
\displaystyle\chi_{\rm sub-GeV}^2
&=&
\min_{\alpha,\beta's}\left[
\frac{\beta_{s1}^2}{\sigma_{\beta s1}^2}
+\frac{\beta_{s2}^2}{\sigma_{\beta s2}^2}\right.
\nonumber\\
&{\ }&+\sum_{j=1}^{10}\left\{
\frac{1}{n_j^{\rm s}(e)}
\left[ \alpha (1-{\beta_{s1} \over 2}+{\beta_{s2} \over 2})N_j^{\rm s}(e)
+ \alpha (1-{\beta_{s1} \over 2}-{\beta_{s2} \over 2})\bar{N}_j^{\rm s}(e)
-n_j^{\rm s}(e)\right]^2
\right.
\nonumber\\
&{\ }&\left.\left.+\frac{1}{n_j^{\rm s}(\mu)}
\left[ 
\alpha (1+{\beta_{s1} \over 2}+{\beta_{s2} \over 2})N_j^{\rm s}(\mu)
+\alpha (1+{\beta_{s1} \over 2}-{\beta_{s2} \over 2})\bar{N}_j^{\rm s}(\mu)
-n_j^{\rm s}(\mu)\right]^2
\right\}\right],\nonumber\\
\displaystyle\chi_{\rm multi-GeV}^2&=&
\min_{\alpha,\beta's}\left[
\frac{\beta_{m1}^2}{\sigma_{\beta m1}^2}
+\frac{\beta_{m2}^2}{\sigma_{\beta m2}^2}\right.
\nonumber\\
&{\ }&+\sum_{j=1}^{10}\left\{
\frac{1}{n_j^{\rm m}(e)}
\left[ \alpha (1-{\beta_{m1} \over 2}+{\beta_{m2} \over 2})N_j^{\rm m}(e)
+ \alpha (1-{\beta_{m1} \over 2}-{\beta_{m2} \over 2})\bar{N}_j^{\rm m}(e)
-n_j^{\rm m}(e)\right]^2
\right.
\nonumber\\
&{\ }&\left.\left.+\frac{1}{n_j^{\rm m}(\mu)}
\left[ 
\alpha (1+{\beta_{m1} \over 2}+{\beta_{m2} \over 2})N_j^{\rm m}(\mu)
+\alpha (1+{\beta_{m1} \over 2}-\frac{\beta_{m2}}{2})\bar{N}_j^{\rm m}(\mu)
-n_j^{\rm m}(\mu)\right]^2
\right\}\right],\nonumber\\
\displaystyle\chi_{\rm upward}^2&=&
\min_{\alpha}\left[
\frac{\alpha^2}{\sigma_{\alpha}^2}
+\sum_{j=1}^{10}
\frac{1}{n_j^{\rm u}(\mu)}
\left[ 
\alpha N_j^{\rm u}(\mu)
-n_j^{\rm u}(\mu)\right]^2\right].
\end{eqnarray}
are $\chi^2$ for the sub-GeV, multi-GeV, and upward going
$\mu$ events, respectively, the summation on $j$ runs
over the ten zenith angle bins for each $\chi^2$, $N_j^a(\alpha)$ and
$n_j^a(\alpha)$ ($a$=s, m, u; $\alpha$=e,$\mu$) stand
for the theoretical predictions and data
for the numbers of the sub-GeV, multi-GeV, and upward going
$\mu$ events,
and it is understood that
$\chi^2$ is minimized with respect to all the normalization
factors $\alpha$, $\beta_s$, $\beta_m$.
We have put $\sigma_{s1}=\sigma_{m1}$ = 0.03,
$\sigma_{s2}=\sigma_{m2}$ = 0.05,
$\sigma_{\alpha}$ = 0.2 and we have assumed that
the overall flux normalization $\alpha$ in
the contained events is a free parameter as in \cite{Ashie:2005ik}, and
we have omitted the other uncertainties, such as the $E_\nu$ spectral index,
the relative normalization between PC and FC and up-down
correlation, etc., for simplicity.

In Eq. (\ref{eqn:chi}) the sum of each $\chi^2$ is optimized
with respect the mixing angle $\theta_{23}$,
the mass squared difference $|\Delta m^2_{32}|$,
the Dirac CP phase $\delta$
and the phase arg($\epsilon_{e\tau}$)
of the parameter $\epsilon_{e\tau}$.
The other oscillation parameters
give little effect on $\chi^2$, so
we have fixed them as $\sin^22\theta_{12}=0.86$,
$\sin^22\theta_{13}=0.1$ and
$\Delta m^2_{21}=7.6\times 10^{-5}$eV$^2$.

\begin{figure}
\vspace*{-7mm}
\includegraphics[scale=0.4]{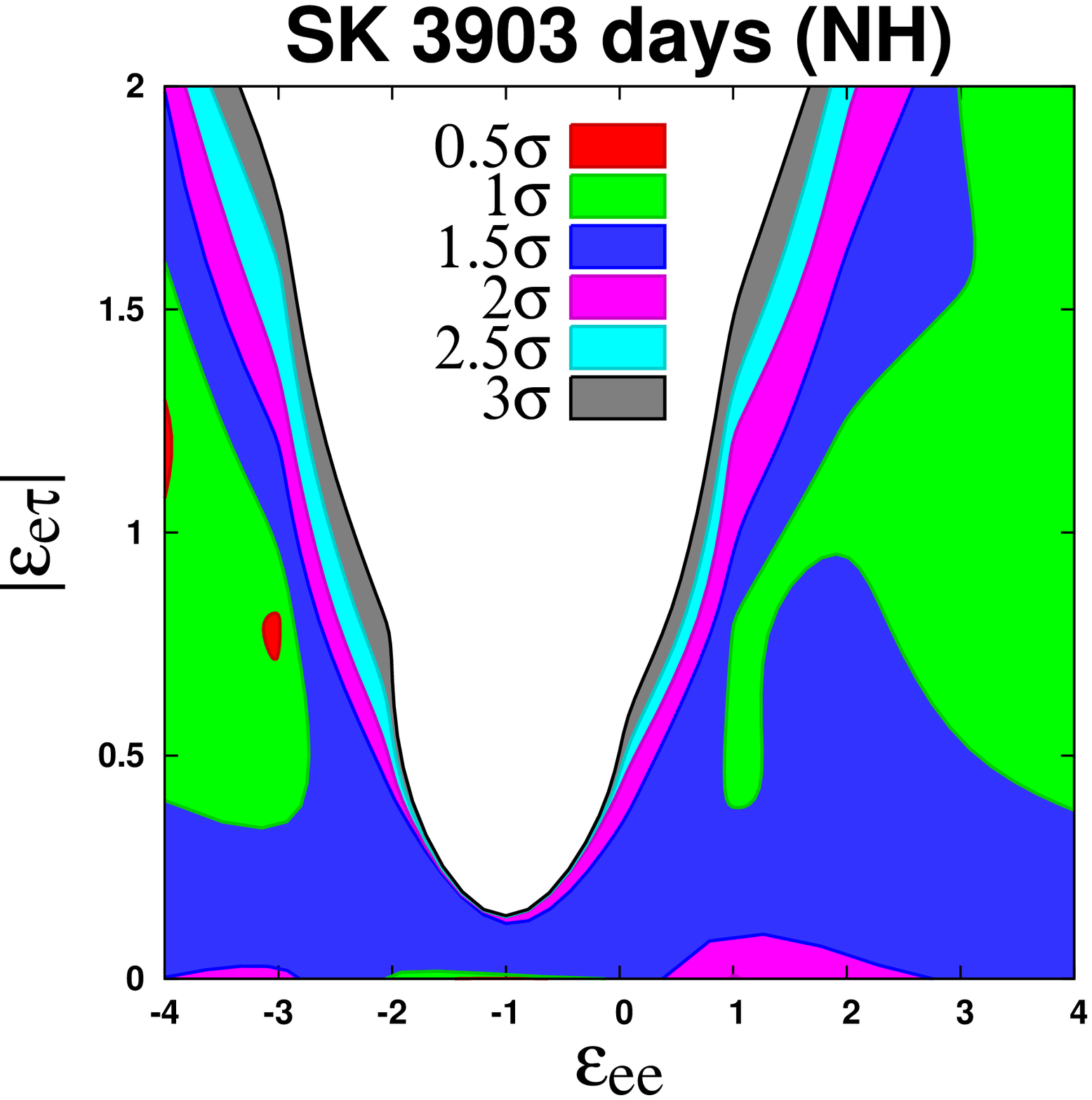}
\includegraphics[scale=0.4]{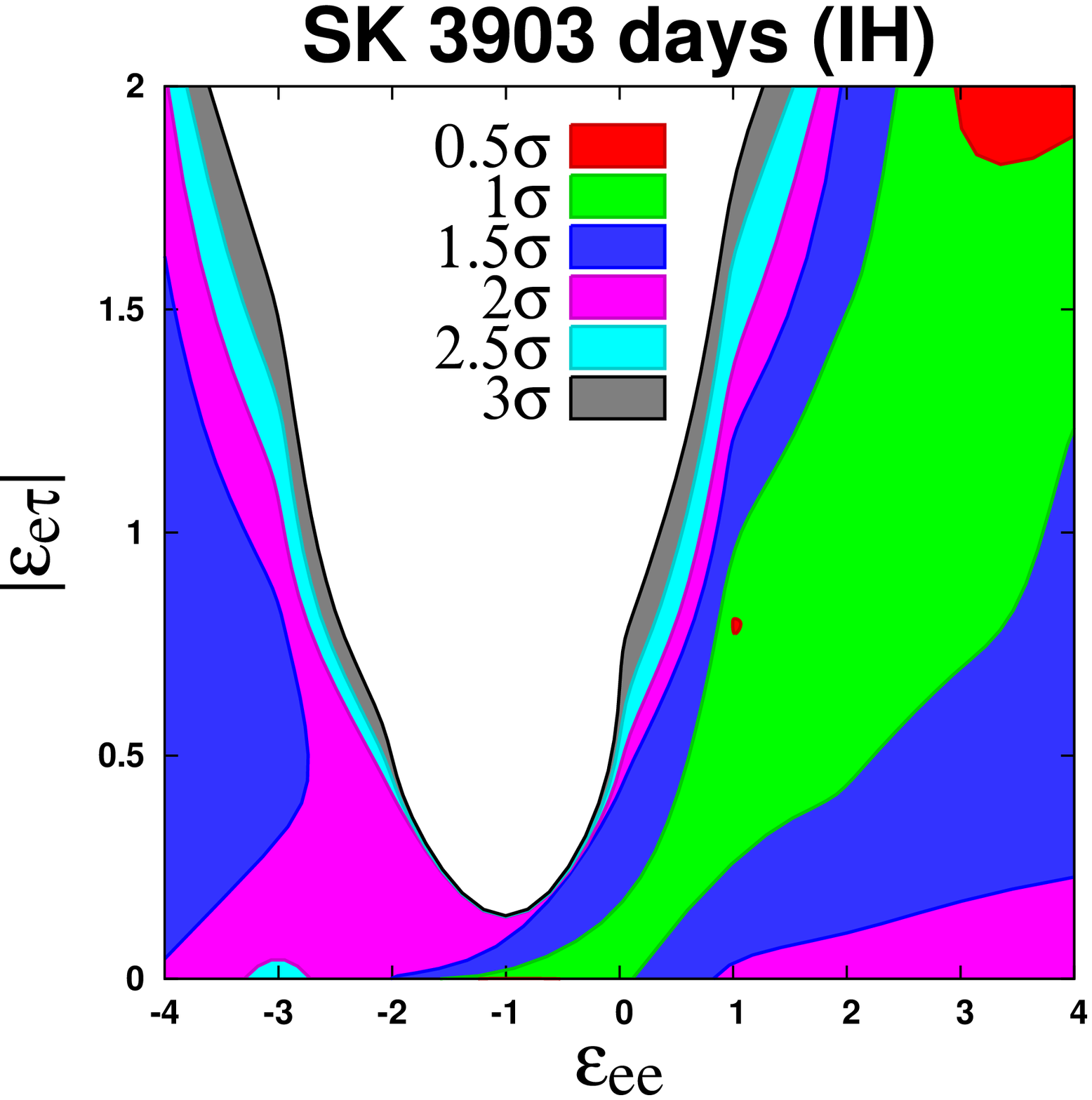}
\vspace*{5mm}
\caption{\label{fig:fig1}
The allowed region in the
($\epsilon_{ee}$, $|\epsilon_{e\tau}|$) plane
of the Superkamiokande atmospheric neutrino data
for 3903 days for the Normal Hierarchy (left panel)
and the Inverted Hierarchy (right panel).
}
\end{figure}

The result for the Superkamiokande data
for 3903 days is given in Fig. \ref{fig:fig1}.
The best-fit point is
$\epsilon_{ee}=-1$, $|\epsilon_{e\tau}|=0$,
and the value of $\chi^2$ at this point is
74.7~(74.5) for 50 degrees of freedom
in the case of the normal (inverted) hierarchy, and
goodness of fit is 2.5 $\sigma$CL in both cases.
The best-fit point is different from the
standard case $\epsilon_{ee}=|\epsilon_{e\tau}|=0$,
and this may be because we have been unable to reproduce
the Monte Carlo
simulation by the Superkamiokande group.
The difference of the value of $\chi^2$ for the
standard case and that for the best-fit point
is $\Delta\chi^2=2.6~(2.0)$ for 2 degrees of freedom
in the case of the normal (inverted) hierarchy,
and its significance is 1.1 (0.9) $\sigma$CL.
So the standard case
is certainly acceptable in our analysis.
From the Fig. \ref{fig:fig1} we observe that
the allowed region for
$|\tan\beta|\equiv|\epsilon_{e\tau}|/|1+\epsilon_{ee}|$
is approximately
$|\tan\beta|\equiv|\epsilon_{e\tau}|/|1+\epsilon_{ee}|\lesssim 0.8$
at 2.5$\sigma$CL.

\begin{figure}
\vspace*{-7mm}
\includegraphics[scale=0.4]{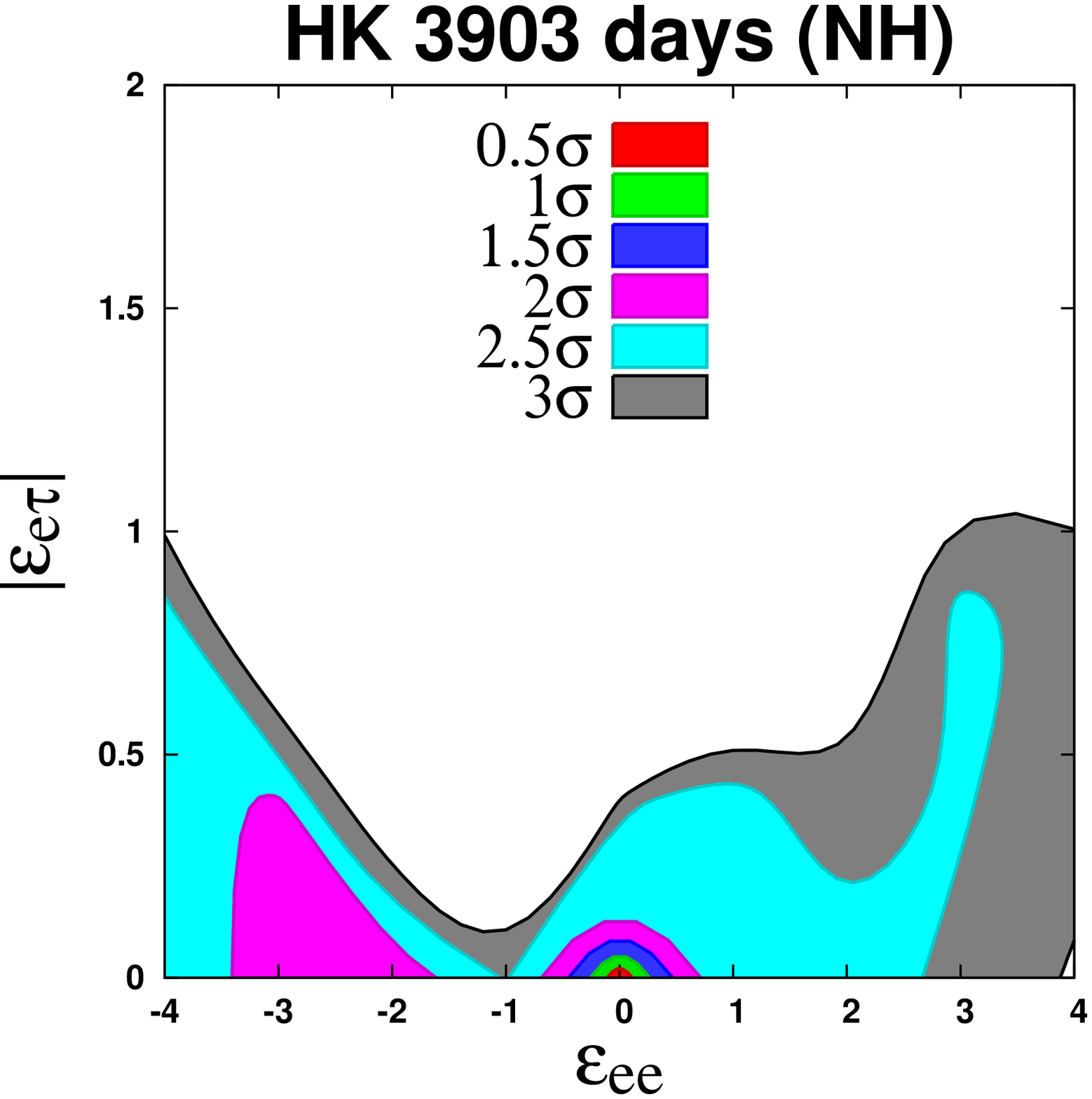}
\includegraphics[scale=0.4]{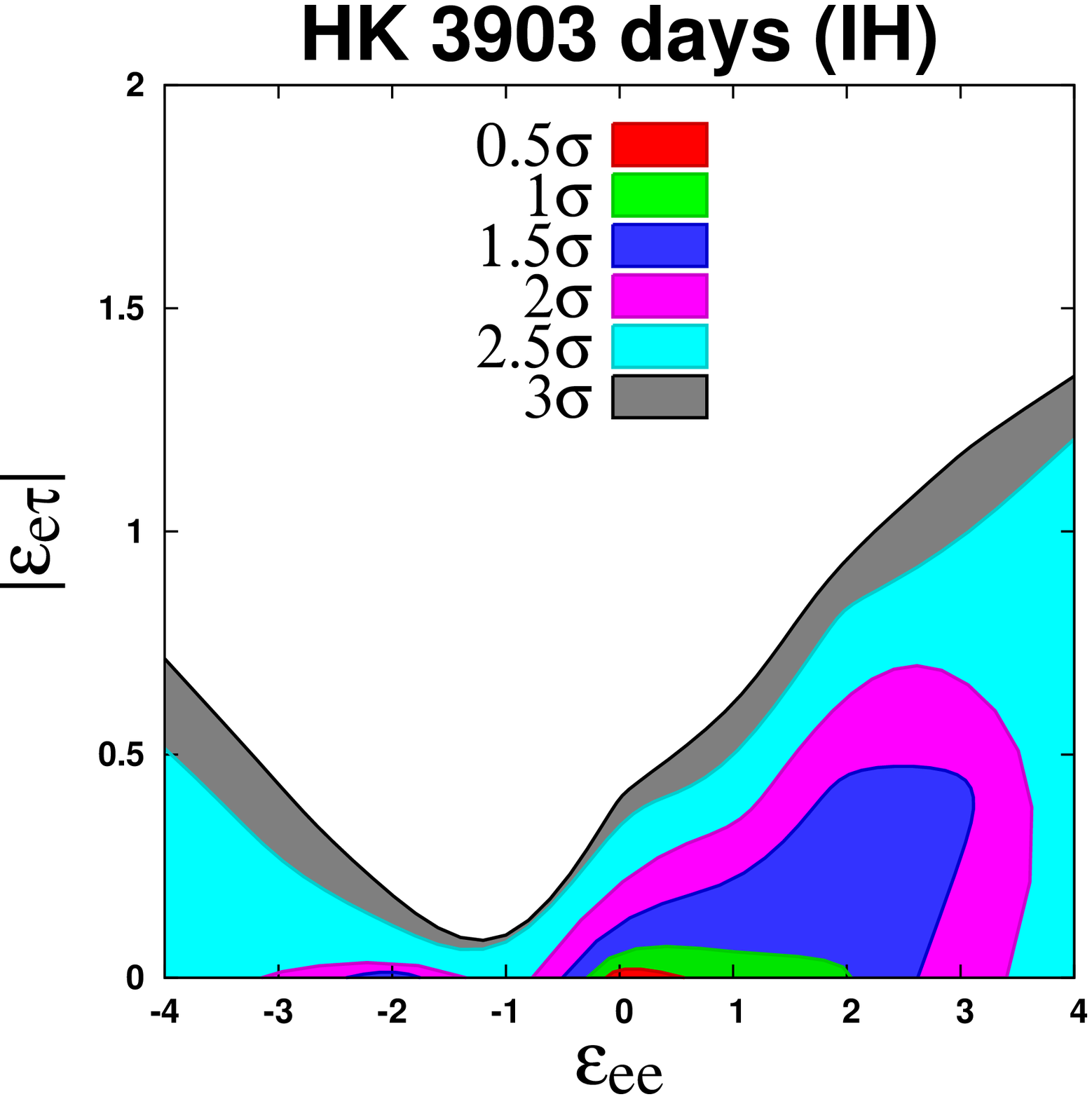}
\vspace*{5mm}
\caption{\label{fig:fig2}
The expected allowed region in the
($\epsilon_{ee}$, $|\epsilon_{e\tau}|$) plane
of the Hyperkamiokande atmospheric neutrino data
for 3903 days for the Normal Hierarchy (left panel)
and the Inverted Hierarchy (right panel).
The best-fit point is the standard case
with $\epsilon_{ee}=|\epsilon_{e\tau}|=0$.
}
\end{figure}

We have also performed the analysis for the
Hyperkamiokande case with the same
period of time, i.e., for 3903 days.
In this case we have assumed that the data
$n_j^a(\alpha)$ ($a$=s, m, u; $\alpha$=e,$\mu$)
are the numbers of events which are expected
from the standard oscillation scenario
with parameters $\theta_{23}=\pi/4$,
$|\Delta m^2_{32}|=2.5\times 10^{-3}$eV$^2$.
We have further assumed for simplicity
that $\delta=0$.
The result for the Hyperkamiokande case
is given in Fig. \ref{fig:fig2}.
In this case
the allowed region for
$|\tan\beta|\equiv|\epsilon_{e\tau}|/|1+\epsilon_{ee}|$
is approximately
$|\tan\beta|\equiv|\epsilon_{e\tau}|/|1+\epsilon_{ee}|\lesssim 0.4$
at 2.5$\sigma$CL.

\section{Conclusion}

In this talk I have shown that the
flavor-dependent neutral current
Non-Standard Interactions in propagation
can be constrained from the atmospheric neutrino
data by Superkamiokande and Hyperkamiokande.
With the ansatz (\ref{matter-nsi1})
for the parameters of the Non-Standard Interactions,
I have shown the allowed region in the
($\epsilon_{ee}$, $|\epsilon_{e\tau}|$) plane.
In the case of the Superkamiokande data for 3903 days,
the allowed region is described by
$|\tan\beta|\equiv|\epsilon_{e\tau}|/|1+\epsilon_{ee}|\lesssim 0.8$
at 2.5$\sigma$CL, while
it is given by
$|\tan\beta|\equiv|\epsilon_{e\tau}|/|1+\epsilon_{ee}|\lesssim 0.4$
at 2.5$\sigma$CL in the Hyperkamiokande
case.

\end{document}